\documentclass[runningheads]{llncs}

\usepackage{graphicx}
\usepackage{graphics}
\usepackage{wrapfig}
\usepackage{float}
\usepackage{textcomp}

\usepackage{graphicx}
\begin{document}
\title{Weakly-supervised Learning for Single-step Quantitative Susceptibility Mapping}
\titlerunning{Weakly-supervised Learning for Single-step QSM}
%
\author{Juan Liu\inst{1}\orcidID{0000-1111-2222-3333} \and
Kevin M. Koch\inst{2,3}\orcidID{1111-2222-3333-4444}}
\authorrunning{J. Liu et al.}
%

\author{{Juan Liu\inst{1,2}} \and
Kevin M. Koch\inst{1,2,3}}
\authorrunning{J. Liu, K.M. Koch}
%
\institute{Center for Imaging Research, Medical College of Wisconsin, Milwaukee, WI, USA \and
Department of Biomedical Engineering, Marquette University and Medical College of Wisconsin, Milwaukee, WI, USA \and
Department of Radiology, Medical College of Wisconsin, Milwaukee, WI, USA\\
\email{juan.liu@marquette.edu}}

\maketitle              
\begin{abstract}

Quantitative susceptibility mapping (QSM) utilizes MRI phase information to estimate tissue magnetic susceptibility. The generation of QSM requires solving ill-posed background field removal (BFR) and field-to-source inversion problems. Because current QSM techniques struggle to generate reliable QSM in clinical contexts, QSM clinical translation is greatly hindered. Recently, deep learning (DL) approaches for QSM reconstruction have shown impressive performance. Due to inherent non-existent ground-truth, these DL techniques use either calculation of susceptibility through multiple orientation sampling (COSMOS) maps or synthetic data for training, which are constrained by the availability and accuracy of COSMOS maps or domain shift when training data and testing data have different domains. To address these limitations, we propose a weakly-supervised single-step QSM reconstruction method, denoted as wTFI, to directly reconstruct QSM from the total field without BFR. wTFI uses the BFR method RESHARP local fields as supervision to perform a multi-task learning of local tissue fields and QSM, and is capable of recovering magnetic susceptibility estimates near the edges of the brain where are eroded in RESHARP and realize whole brain QSM estimation. Quantitative and qualitative evaluation shows that wTFI can generate high-quality local field and susceptibility maps in a variety of neuroimaging contexts. 

\end{abstract}

\section{Introduction}
Quantitative susceptibility mapping (QSM) can estimate tissue magnetic susceptibility values from MRI Larmor frequency sensitive phase images to provide novel image contrast \cite{wang2015quantitative}. To date, all QSM methods rely on a dipolar convolution that relates susceptibility sources to induced Larmor frequency offsets \cite{salomir2003fast,marques2005application}, which is expressed in the k-space as bellow. 

\begin{equation} 
B(\vec k) = X (\vec k) \cdot {D}(\vec k); D(\vec k) =  \frac{1}{3} - \frac{{k_z^2}}{{{k_x^2 + k_y^2 + k_z^2}}}
\end{equation}

where $B(\vec k)$ is the induced magnetic perturbation along the main magnetic field $B_{0}$ direction, $X(\vec k)$ is the susceptibility distribution in the k space, $D(\vec k)$ is the dipole kernel.

The generation of QSM requires solving two challenging ill-posed problems - (1) removal of background field contributions from sources outside the interest, (2) field-to-source inversion by solving the dipole deconvolution. Though existing BFR algorithms demonstrate excellent performance, they have several limitations, including volume erosion, inaccurate BFR close to volume boundary, and residual background leakage \cite{schweser2017illustrated}. Incorrect BFR often introduces erroneous local field outputs and subsequently affects susceptibility quantification.

The field-to-source inversion faces technical challenges due to the singularities of the dipole kernel. Calculation of susceptibility through multiple orientation sampling (COSMOS) \cite{liu2009calculation} remains the empirical gold-standard of QSM, as the additional field data sufficiently improves the conditioning of this ill-posed inversion. However, multi-orientation data acquisition is time consuming and clinically infeasible. Single-orientation QSM is preferred which is typically computed by either thresholding of the convolution operator \cite{shmueli2009magnetic,wharton2010susceptibility} or sophisticated regularization methods \cite{de2008quantitative,de2010quantitative,liu2011morphology,bilgic2014fast}. In addition, several single-step QSM methods \cite{chatnuntawech2017single,liu2017preconditioned,sun2018whole} have been proposed to directly estimate QSM from the total field (combined BFR and dipole inversion) to prevent potential error propagation across successive operations. However, existing QSM techniques still struggle to generate reliable QSM estimates in clinical contexts, which greatly hinders QSM clinical translation.

Recently, several deep learning (DL) QSM techniques have been proposed. For dipole inversion, QSMnet \cite{yoon2018quantitative} and QSMGAN \cite{chen2019qsmgan} utilized COSMOS estimates as QSM labels, while DeepQSM \cite{bollmann2019deepqsm} was trained using purely synthetic data. AutoQSM \cite{wei2019learning} utilized STAR-QSM \cite{wei2015streaking} estimates after SMV method \cite{wu2012whole} for BFR as QSM labels for single-step QSM learning. Though these techniques demonstrate promising performance, they have several limitations. Due to the lack of a ground-truth reference, these methods usually used COSMOS maps or synthetic data for training. However, acquiring large amounts of COSMOS data is not only expensive but also time consuming. In addition, COSMOS neglects tissue susceptibility anisotropy \cite{liu2010susceptibility} and contains errors from BFR and image registration procedures, which compromise its value as a training label. Though synthetic data generated from the physical model provides a reliable and cost-effective way for training, the generalization needs to be addressed. In autoQSM, the robustness and accuracy of STAR-QSM could affect the performance of autoQSM. Moreover, for DL approaches for dipole inversion only, their performance still is affected by the performance of BFR methods.  

Here, we propose a weakly-supervised approach for single-step QSM, denoted as wTFI. wTFI utilizes RESHARP \cite{sun2014background} local fields as supervision for a multi-task learning of local tissue fields and QSM. For QSM quantitative evaluation, 9 multi-orientation datasets are utilized with comparison to TKD \cite{shmueli2009magnetic}, MEDI \cite{liu2012morphology}, and STAR-QSM, using COSMOS result as a reference. In addition, the local tissue fields of wTFI are qualitatively compared with BFR methods, SHARP \cite{schweser2010differentiation}, RESHARP \cite{sun2014background}, PDF \cite{liu2011novel}, and LBV \cite{zhou2014background}. More qualitative analysis is performed on single-orientation datasets and clinical datasets. 

\section{Method}

\begin{figure}[H]
\vspace{-10pt}
\begin{center}
\includegraphics[width=\textwidth]{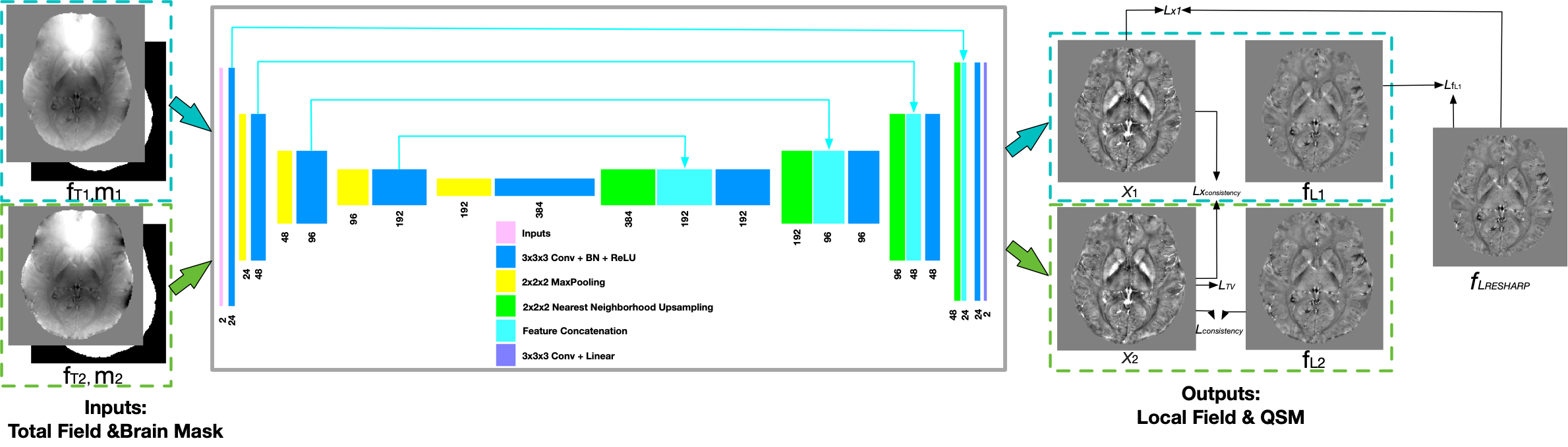}
\vspace{-5pt}
\caption{Network structure of wTFI. It has an encoder-decoder structure with 9 convolutional layers (kernel size 3x3x3, same padding), 9 batch normalization layers, 9 ReLU layers, 4 max pooling layers (pooling size 2x2x2, strides 2x2x2), 4 nearest-neighbor unsampling layers (size 2x2x2), 4 feature concatenations, and 2 convolutional layer (kernel size 3x3x3, linear activation).} 
\label{figCNN}
\end{center}
\vspace{-30pt}
\end{figure}

WTFI was trained using a 3D convolutional neural network (CNN) with an encoder-decoder structure, as shown in Fig.\ref{figCNN}. Since RESHARP produces more accurate local fields when compared with other BFR methods such as SHARP, PDF, and LBV. Local fields of RESHARP were utilized as supervision in the training paradigm. However, RESHARP suffered from brain erosion at the expense of losing information of local field and thus QSM measures in these regions. To address this problem, wTFI utilized a domain adaption technique to recover the lost information at brain edges for whole brain QSM.

During training, wTFI took 4 inputs (2 groups of total fields and brain masks) to get 4 outputs (2 groups of local fields and susceptibility maps). For the total field $f_{T1}$ and brain mask $m_{1}$, the corresponding outputs were $\chi _{1}$ and $f_{L1}$, and $f_{T2}$ and $m_{2}$ with outputs $\chi _{2}$ and $f_{L2}$. Define $m_{2}$ as brain mask obtained using human brain extraction tools from the MR magnitude data; $m_{1}$ as the eroded brain region after RESHARP, which has a smaller region than $m_{2}$ ($m_{2}>m_{1}$); $f_{L_{RESHARP}}$ as RESHARP local field; $f_{T}$ as the total field estimated from phase images; $f_{T1}$ as the total field in brain region $m_{1}$, $f_{T1}=m_{1} f_{T}$; $f_{T2}$ as the total field in brain region $m_{2}$, $f_{T2}=m_{2} f_{T}$.

The loss function consisted of five terms. The first loss term, $L_{\chi _{1}}$, was imposed on $\chi _{1}$. Leveraging the local field results from RESHARP $f_{L_{RESHARP}}$ as a weak supervision, $\chi _{1}$ convoluted with the dipole kernel $d$ should satisfy the well-established QSM inversion physical model. 

\begin{equation}
L_{\chi _{1}} = \left \|m_{1}W(e^{f_{L_{RESHARP}}} - e^{d \ast \chi _{1}})\right \|_{2}
\end{equation}

where $W$ is a data-weighting factor which can be the magnitude image or noise weight matrix, $\ast$ is the convolution operator. Since noise is unknown and spatially variant in the local field measurements, the nonlinear dipole convolution data consistency loss was used to get more robust QSM estimates as conventional QSM methods \cite{liu2013nonlinear,polak2020nonlinear}. 

Next, we included $L_{f_{L1}}$ on local field output $f_{L1}$,
\begin{equation}
L_{f_{L1}} = \left \|m_{1}(f_{L1}-f_{L_{RESHARP}})\right \|_{2}
\end{equation}

Next, a data consistency loss was imposed on $\chi_{2}$ and $f_{L2}$. 
\begin{equation}
L_{consistency} = \left \|m_{2}W(e^{f_{L2}} - e^{d\ast \chi_{2}})\right \|_{2}
\end{equation}

Furthermore, define the susceptibility consistency loss between $\chi_{1}$ and $\chi_{2}$ inside $m_{1}$.
\begin{equation}
L_{\chi_{consistency}} = \left \|m_{1}(\chi_{1}-\chi_{2})\right \|_{2}
\end{equation}

The total variation (TV) loss $L_{TV}$ on $\chi_{2}$ serves as regularization for edge preserving and denoising on the QSM.

\begin{equation}
L_{TV} = \left \|G_{x}(\chi_{2})\right \|_{1} + \left \|G_{y}(\chi_{2})\right \|_{1} + \left \|G_{z}(\chi_{2})\right \|_{1}
\end{equation}

where $G_{x}$, $G_{y}$, and $G_{z}$ are gradient operator in x, y, z directions. 

\begin{equation}
L_{total} = L_{\chi _{1}} + \lambda _{1}L_{f_{L1}} + \lambda _{2}L_{consistency} +  \lambda _{3} L_{\chi_{consistency}} +  \lambda _{4}L_{TV}
\end{equation}

The total loss $L_{total}$ was the weighted sum of 5 loss functions. $\lambda _{1}$, $\lambda _{2}$, $\lambda _{3}$, and $\lambda _{4}$ were loss weights.

After training, the trained DL model only took the whole brain total field $f_{T2}$ and brain mask $m_{2}$ to get the local field and susceptibility map.   

\section{Experiments}
\textbf{Multi-orientation QSM Data} 
9 QSM datasets were acquired using 5 head orientations and a 3D GRE scan with voxel size 1x1x1 mm$^3$ from 3T MRI scanners, shared by Dr. Jongho Lee \cite{yoon2018quantitative}. QSM data processing was implemented as below, offline GRAPPA \cite{griswold2002generalized} reconstruction used to reconstruct magnitude and phase images from saved k-space data, BET (FSL, FMRIB, Oxford, UK) \cite{smith2002fast} for brain mask extraction, the Laplacian method \cite{li2011quantitative} for phase unwrapping, and RESHARP \cite{sun2014background} with spherical mean radius 4mm for BFR. COSMOS results were calculated using the 5 orientation data with image registration using FLIRT (FSL, FMRIB, Oxford, UK) \cite{jenkinson2002improved,jenkinson2001global}. TKD, MEDI, STAR-QSM were performed to get QSM estimates at normal head position.

For wTFI training, leave-one-out cross validation was used. For each dataset, a total of 40 scans from other 8 datasets were used for network training. wTFI was trained using patch-based neural network with patch size 96x96x96. Around 2000 patch pairs of total fields, brain masks, RESHARP local fields with patch size 96x96x96 with an overlapping of 16.6 percent between adjacent patches were cropped. After training, the trained DL model took the whole brain total field and brain mask of the leave-one dataset to get the local field and QSM. 

QSM images at the eroded brain region were quantitatively compared with respect to COSMOS maps, using peak signal-to-noise ratio (pSNR), normalized root mean squared error (NRMSE), high-frequency error norm (HFEN), and structure similarity index (SSIM). Due to no ground-truth, the local fields of SHARP, RESHARP, PDF, LBV, and wTFI were qualitatively compared.

\textbf{Single Orientation QSM Data} 
200 QSM datasets were collected on a 3T MRI scanner (GE Healthcare MR750) from a susceptibility-weighted software application (SWAN, GE Healthcare). The data acquisition parameters were as follows: in-plane data matrix 320x256, FOV 24 cm, voxel size 0.5x0.5x2.0 mm$^3$, 4 TEs [10.4, 17.4, 24.4, 31.4] ms, TR 58.6 ms, and total acquisition time 4 minutes.

Complex multi-echo images were reconstructed from saved k-space data. The brain masks were obtained using the SPM tool \cite{brett2002region}. After BFR using the RESHARP with spherical mean radius 4mm, susceptibility inversion was performed using TKD, STAR-QSM, and MEDI. For wTFI training, 8000 patches of total field map, brain mask and RESHARP results from 100 datasets with patch size 128x128x64 were used for training.

\textbf{Clinical Data} 150 clinical data were acquired using susceptibility-weighted angiography (SWAN, GE Healthcare, Waukesha WI) on a 3T MRI scanner (GE Healthcare MR750) with the following data acquisition parameters: in-plane data matrix 288x224, FOV 22 cm, slice thickness 3 mm, first TE 12.6 ms, echo spacing 4.1 ms, 7 echoes, TR 39.7 ms, pixel bandwidth 244 Hz, and total acquisition time of about 2 minutes. 

Complex multi-echo images were reconstructed from raw k-space data. The brain masks were obtained using the SPM tool \cite{brett2002region}. RESHARP method with spherical mean radius 6mm were used for BFR. For wTFI training, about 2000 patch pairs of total fields, brain masks and local fields from 100 datasets with patch size 128x128x64 were used. 

\section{Experimental Results}

\begin{figure}[H]
\begin{center}
\vspace{-35pt}
\includegraphics[width=0.9\textwidth]{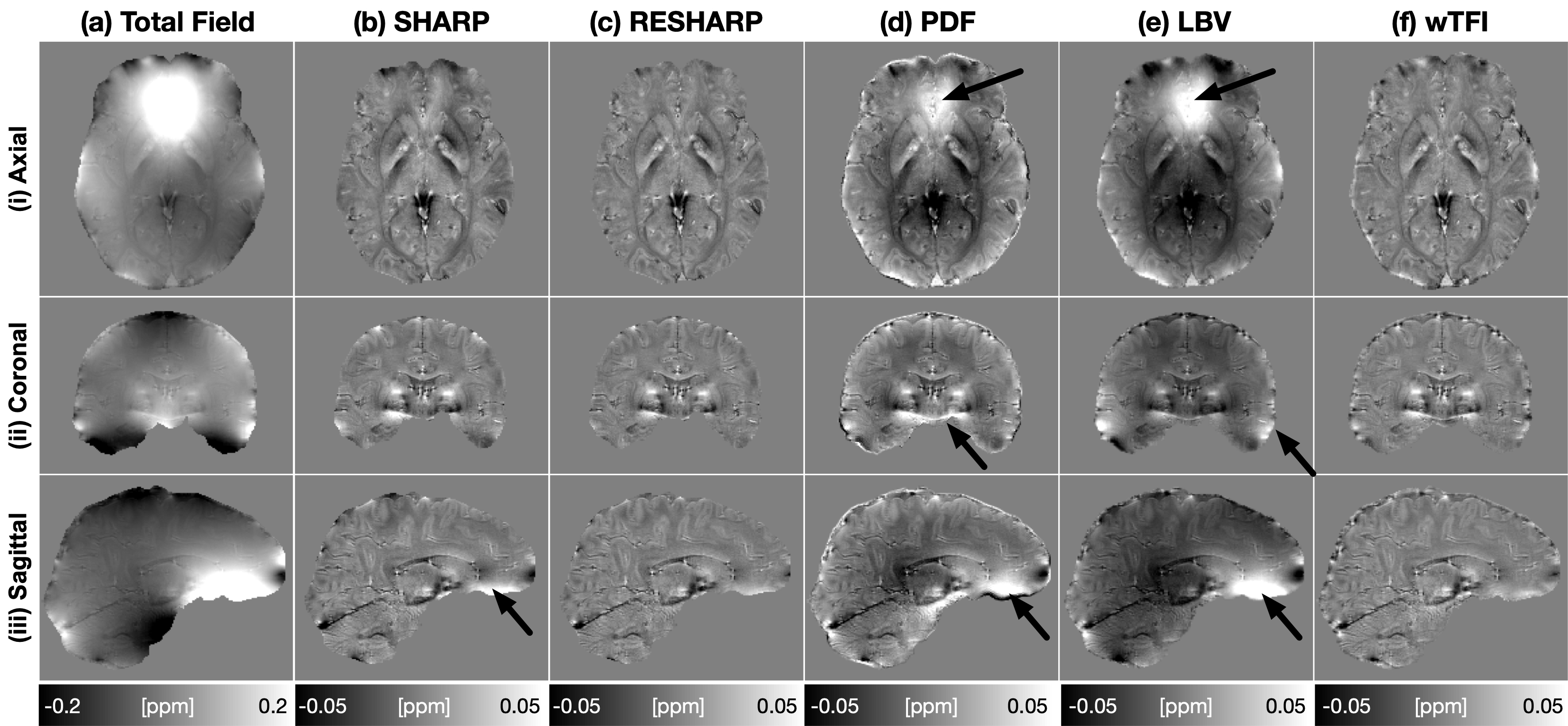}
\vspace{-5pt}
\caption{Total fields and local fields of a multi-orientation dataset.} 
\label{fig_7_RDF}
\vspace{-20pt}
\end{center}
\end{figure}

\begin{figure}[H]
\begin{center}
\vspace{-20pt}
\includegraphics[width=0.9\textwidth]{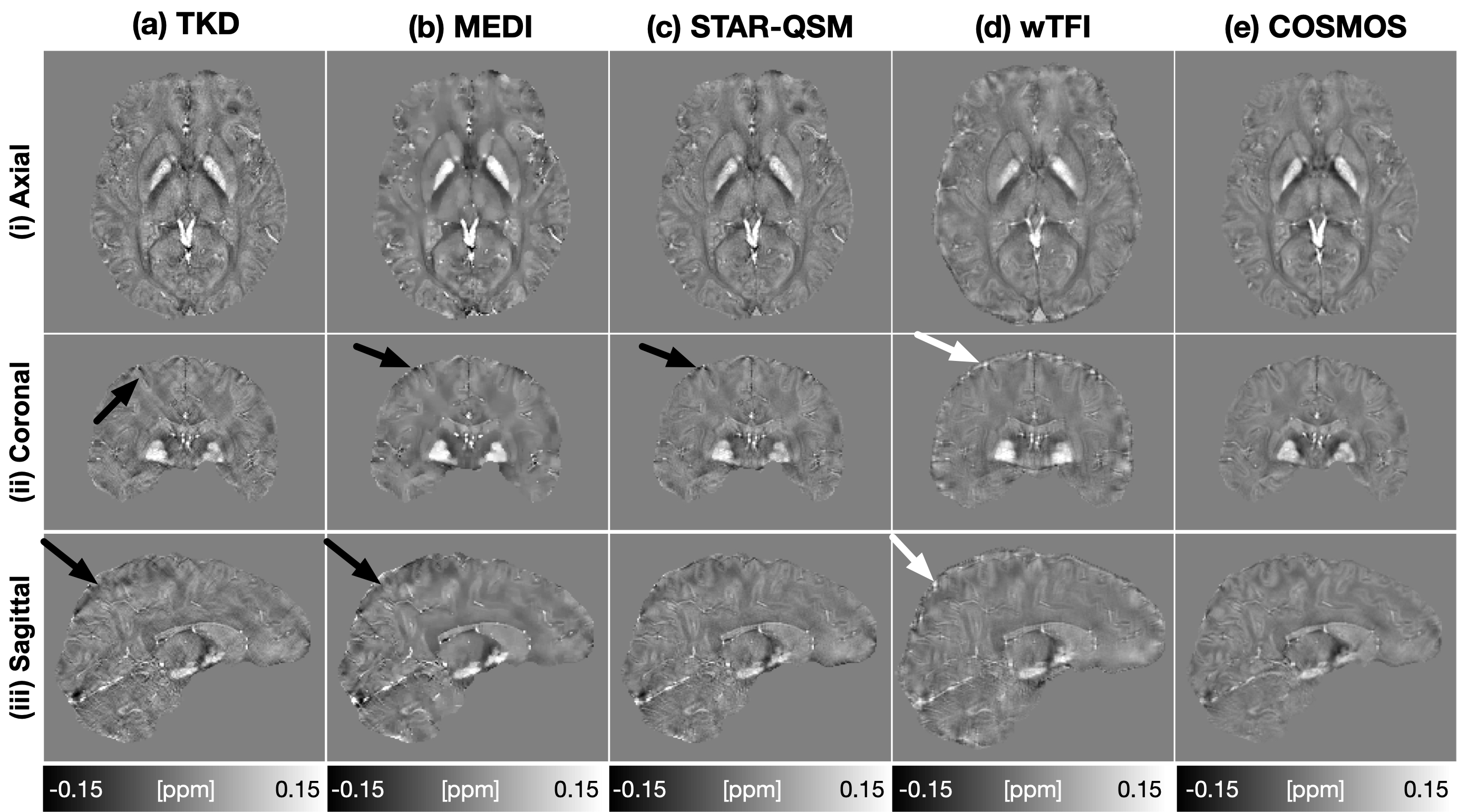}
\vspace{-10pt}
\caption{Comparison of QSM of a multi-orientation dataset.} 
\label{fig_7_QSM}
\vspace{-25pt}
\end{center}
\end{figure}

\textbf{Multi-orientation Data} 
Table.\ref{table_1} summarized the quantitative metrics from 4 QSM methods on 9 datasets with COSMOS map as a reference. Compared to TKD, STAR-QSM, and MEDI, wTFI achieved the best metric scores in pSNR, NRMSE, and second in HFEN. Fig.\ref{fig_7_RDF} showed the total fields and local fields from a representative dataset. Residual background fields showed up in SHARP (b), PDF (d), and LBV (e) results. SHARP (b) and RESHARP (c) suffered from brain erosion. PDF and LBV results showed strong shading artifacts and erroneous BFR. wTFI (f) produced RESHARP-like local fields and preserved the whole brain without erosion. Fig.\ref{fig_7_QSM} displayed the QSM images. TKD (a) and MEDI (b) results showed streaking artifacts (a-b, black arrows). MEDI images lost details due to oversmoothing. STAR-QSM (c) images showed good quality with slight image artifacts. wTFI (d) produced high quality QSM and recovered the susceptibility information of brain edges (d, white arrows).

\begin{table}[!ht]
\centering
\vspace{-10pt}
\caption{\label{table_1} Means and standard deviations of quantitative  performance metrics from 4 reconstruction methods on 9 datasets.}
\vspace{0in}
\begin{tabular}{cccccc}
\hline
\multicolumn{1}{|c}{} & \multicolumn{1}{|c}{pSNR (dB)} & \multicolumn{1}{|c}{NRMSE ($\%$)} &\multicolumn{1}{|c}{HFEN ($\%$)} &\multicolumn{1}{|c|}{SSIM (0-1)}\\
\hline 
\multicolumn{1}{|c}{TKD} & \multicolumn{1}{|c}{$43.4\pm0.5$} & \multicolumn{1}{|c}{$91.4\pm6.7$} & \multicolumn{1}{|c}{$72.9\pm6.6$} & \multicolumn{1}{|c|}{$0.831\pm0.016$}\\
\hline
\multicolumn{1}{|c}{MEDI} & \multicolumn{1}{|c}{$41.5\pm0.6$} & \multicolumn{1}{|c}{$113.8\pm7.6$} & \multicolumn{1}{|c}{$100.4\pm9.1$} & \multicolumn{1}{|c|}{\textbf{0.902$\pm$0.016}}\\
\hline
\multicolumn{1}{|c}{STAR-QSM} & \multicolumn{1}{|c}{$45.1\pm0.5$} & \multicolumn{1}{|c}{$75.4\pm5.4$} & \multicolumn{1}{|c}{\textbf{61.7$\pm$4.7}} & \multicolumn{1}{|c|}{$0.876\pm0.016$}\\
\hline
\multicolumn{1}{|c}{wTFI} & \multicolumn{1}{|c}{\textbf{45.3$\pm$0.5}} & \multicolumn{1}{|c}{\textbf{73.8$\pm$4.2}} & \multicolumn{1}{|c}{{66.2$\pm$3.4}} & \multicolumn{1}{|c|}{$0.870\pm0.017$}\\
\hline 
\end{tabular}
\begin{flushleft}
\end{flushleft}
\vspace{-20pt}
\end{table}

\begin{figure}[H]
\vspace{0pt}
\begin{center}
\includegraphics[width=0.9\textwidth]{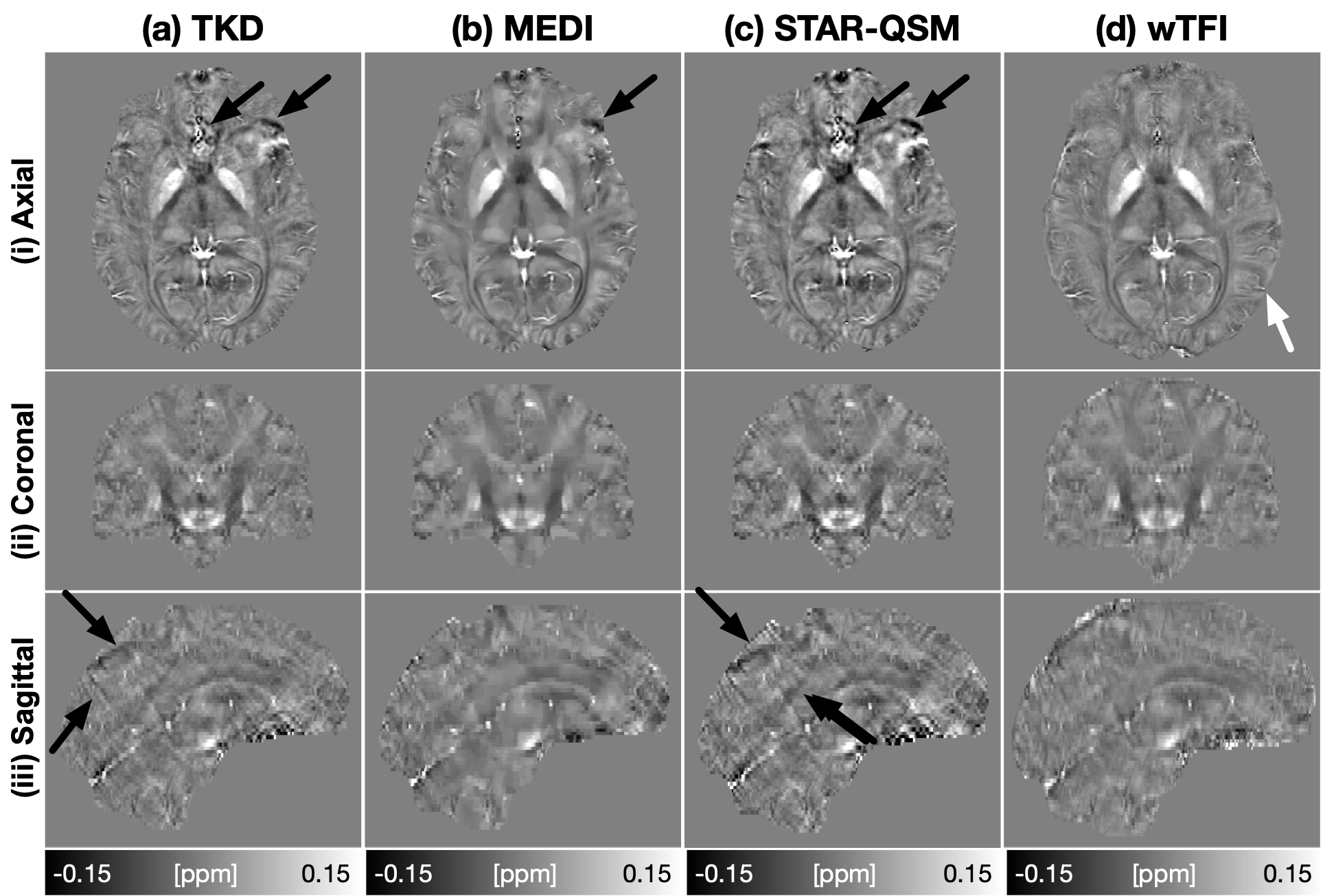}
\caption{Comparison of QSM estimates of a single-orientation dataset.}
\label{fig_H2H}
\vspace{-25pt}
\end{center}
\end{figure}

\textbf{Single-orientation Data} 
Fig.\ref{fig_H2H} displayed the QSM estimates of a single-orientation dataset. Visually, wTFI (d) outperformed TKD (a), MEDI (b) and STAR-QSM (c) with invisible streaking and shading artifacts. In addition, WTFI was capable of recovering the susceptibility information at brain boundaries.  

\begin{figure}[H]
\vspace{-20pt}
\begin{center}
\includegraphics[width=\textwidth]{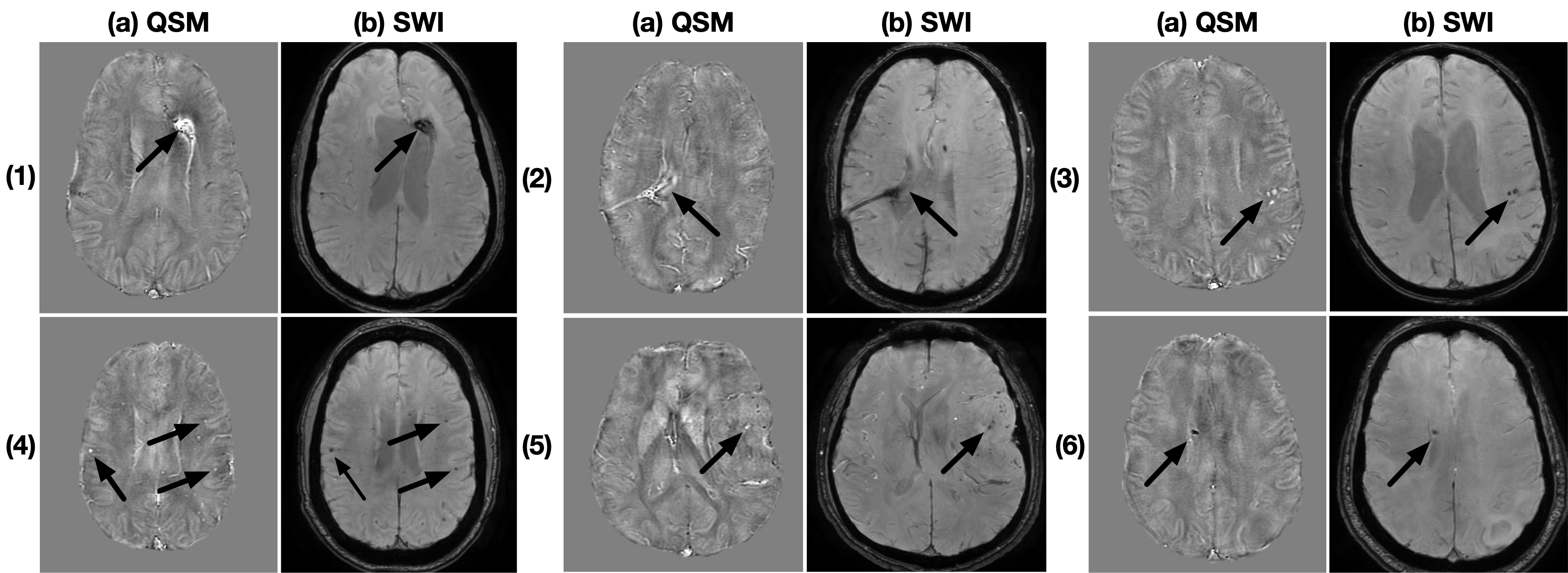}
\vspace{-10pt}
\caption{QSM and SWI images from 6 clinical patients. (1) a 50-year-old patient with central nervous system (CNS) Lymphoma, (2) a 30-year-old patient with encephalopathic with headache and meningitis, (3) a 80-year-old patient with Acute changes in executive functioning and metastatic bladder cancer, (4) a 32-year-old patient with metastatic non-small cell lung cancer to the brain, (5) a 48-year-old patient with left cerebral convexity meningioma, (6) a 51-year-old patient with tuberous sclerosis.} 
\label{fig_Clinical}
\vspace{-20pt}
\end{center}
\end{figure}

\textbf{Clinical Data} 
Fig.\ref{fig_Clinical} displayed SWI images and susceptibility maps calculated using wTFI from 6 clinical patients. wTFI results clearly showed the brain hemorrhage, microbleeds, calcification, and vessel malformation.

\textbf{Ablation Study}
The ablation study was to investigate the neural network design. 3 neural networks were compared, (1) wTFI, (2) wTFI$_{m}$ only trained with inputs $f_{T1}$ and $m_{1}$. (3) STAR-Net, which used $f_{T1}$ and $m_{1}$ as inputs and STAR-QSM maps as QSM labels for single-step QSM reconstruction.

\begin{figure}[H]
\vspace{0pt}
\begin{center}
\includegraphics[width=0.9\textwidth]{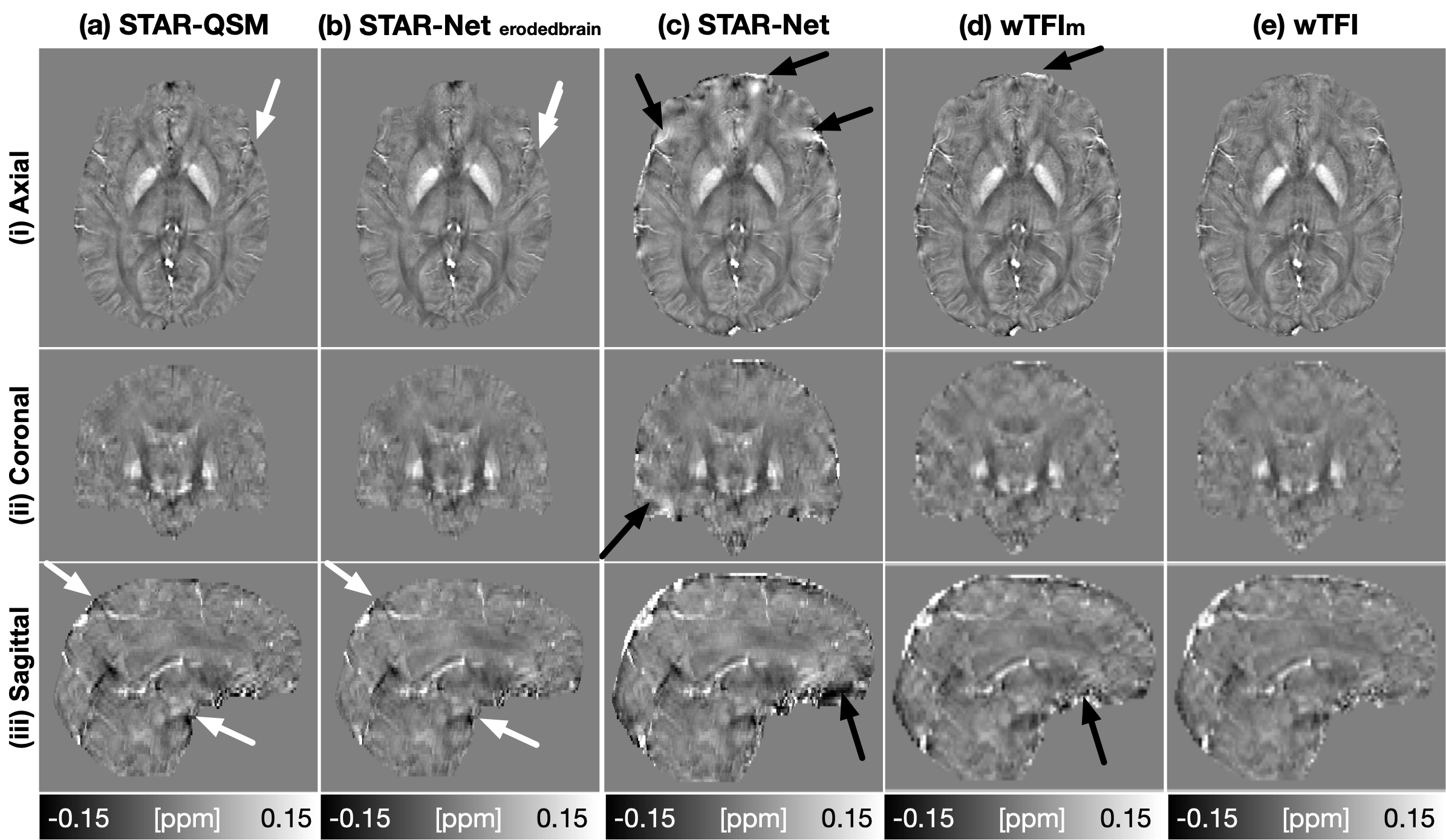}
\vspace{-5pt}
\caption{Comparison of QSM of a single-orientation dataset.} 
\label{fig_Ablation}
\vspace{-25pt}
\end{center}
\end{figure}

Fig.\ref{fig_Ablation} showed QSM estimates of a single-orientation dataset. STAR-Net in eroded brain (b) showed STAR-QSM-like quality, but less black shadings (white arrows). Whole brain STAR-Net (c) and wTFI$_{m}$ (d) showed larger errors close to brain boundary when applying to whole brain $f_{T2}$ and $m_{2}$ for QSM estimates (black arrows). This indicated the domain shift caused susceptibility quantification errors. Visual assessment showed that wTFI (e) produced better whole brain susceptibility maps. 

\section{Discussion and Conclusion}
In this work, a weakly-supervised DL method for single-step QSM was proposed. From quantitative evaluation, wTFI achieved high metric scores. However, COSMOS as reference has limitations aforementioned. Based on visual assessment, wTFI outperforms conventional background field removal and dipole inversion methods. Most importantly, wTFI is capable of recovering the QSM information close to brain edges.  

The proposed wTFI for QSM reconstruction approach has several advantages. First, wTFI employs RESHARP results as weak supervision and does not require QSM labels. Second, wTFI performs a multi-task learning, which estimates local fields and susceptibility map. Third, wTFI is able to recover magnetic susceptibility of anatomical structures near the edges of the brain. wTFI has a limitation that the wTFI performance is affected by RESHARP.    

\section*{Acknowledgement}
We thank Professor Jongho Lee for sharing the multi-orientation QSM datasets.

%
%
\bibliographystyle{splncs04}
\bibliography{reference}
\end{document}